\documentstyle[preprint,prd,aps,epsfig,floats]{revtex}
\tighten
\let\jnfont=\rm
\def\NPB#1,{{\jnfont Nucl.\ Phys.\ B }{\bf #1},}
\def\PLB#1,{{\jnfont Phys.\ Lett.\ B }{\bf #1},}
\def\PRD#1,{{\jnfont Phys.\ Rev.\ D }{\bf #1},}
\def\PRL#1,{{\jnfont Phys.\ Rev.\ Lett.\ }{\bf #1},}
\def\MPLA#1,{{\jnfont Mod.\ Phys.\ Lett.\ A }{\bf #1},}
\def\JPG#1,{{\jnfont J.\ Phys.\ G}{\bf #1},}
\def\CTP#1,{{\jnfont Commun.\ Theor.\ Phys.\ }{\bf #1},}
\def\Rv{\not{\hbox{\kern-4pt $R$}}}
\def\Bv{\not{\hbox{\kern-4pt $B$}}}
\def\Lv{\not{\hbox{\kern-4pt $L$}}}
\newcommand{\gsim}{\mathrel{\lower4pt\hbox{$\sim$}}
\hskip-12.5pt\raise1.6pt\hbox{$>$}\;}
\newcommand{\lsim}{\mathrel{\lower4pt\hbox{$\sim$}}
\hskip-12.5pt\raise1.6pt\hbox{$<$}\;}
\begin{document}

\preprint{
\hfill$\vcenter{ 
 \hbox{\bf MADPH-01-1212} 
{\hbox{\bf hep-ph/0102037} }
}$ 
}

\title{\ \\[5mm] Top-quark rare decay $t\to c h$ in R-parity-violating SUSY}

\author{\ \\[2mm] G.~Eilam$^1$, A. Gemintern$^1$,
T.~Han$^2$, J.~M.~Yang$^3$ and X.~Zhang$^4$}

\address{\ \\[2mm]   
$^1$ {\it Physics Department, Technion, 32000 Haifa, Israel}\\[2mm]
$^2$ {\it Department of Physics, University of Wisconsin, 
               Madison, WI 53706, USA}\\[2mm]
$^3$ {\it Institute of Theoretical Physics, Academia Sinica,
          Beijing 100080, China}\\[2mm]
$^4$ {\it Institute of High Energy Physics, Academia Sinica, 
         Beijing 100039, China} }

\maketitle

\begin{abstract}
The flavor-changing top-quark decay $t\to c h$, where $h$ is the lightest 
CP-even Higgs boson in the minimal supersymmetric standard model, 
is examined in the R-parity-violating supersymmetric model.  
Within the existing bounds on the relevant R-parity-violating 
couplings, the branching fraction for $t\to c h$ can be as large
as about $10^{-5}$ in some part of the parameter space.
\end{abstract}


%\begin{center} {\Large 1. Introduction }\end{center}

The study of heavy-particle decays via 
flavor-changing neutral-currents (FCNC)
has been playing an important role in testing the standard model (SM) and 
probing new physics beyond the SM.  
As the heaviest elementary particle in the SM
with a mass at the electroweak scale, the top quark is more
likely to be sensitive to new physics.
Kinematically it is accessible to many FCNC decay modes, such as  
$t\to cV$ ($V=\gamma, Z, g$) and  $t\to c h$, where $h$ is a Higgs
boson. In the SM these FCNC decay modes are highly suppressed 
by GIM mechanism, with branching fractions typically
of $10^{-13}-10^{-10}$ \cite{tcvh_sm,tch_sm}, which 
are too small to be detectable at collider experiments. On the other hand,
observation of any of such FCNC top-quark decays would be robust 
evidence for new physics \cite{tcv_mc,tcvmore,tch_mc,mIII}.

Top quarks will be copiously produced at the next generation of hadron
colliders. At the upgraded Fermilab Tevatron with an integrated
luminosity of 10 fb$^{-1}$, there will be about $8\times 10^4$ top
quarks produced, while there will be about 100 times more at the LHC
with the same luminosity.
With such large data samples, good sensitivities may be reached for
searching for the rare decay channels $t\to cV$ \cite{tcv_mc}, 
and for studying other related processes at hadron colliders \cite{tcvmore}.
A more recent study showed that the channel $t\to ch$ could
also be detectable \cite{tch_mc}, reaching a sensitivity level 
for the branching fraction $Br(t\to ch)\sim 5\times 10^{-5}$ 
at the LHC and a few percent at the Tevatron. 
While these high detection sensitivities are still far above 
the SM expectation for the rare decay channels \cite{tcvh_sm,tch_sm},
in many scenarios beyond the SM the branching fractions of
these FCNC top-quark decays could be significantly enhanced
\cite{tcvh_sm,mIII,tcv_mssm,tch_mssm,tcv_rv,tcsnu,tcv_tc}. 

In the minimal supersymmetric (SUSY) 
standard model (MSSM) \cite{mssm} with R-parity conservation, 
it was shown \cite{tch_mssm} that the possibility for observing the decay 
channel  $t\to ch$ could be greatly enhanced 
(here $h$ is the lightest CP-even Higgs boson). 
Kinematically this decay mode is always allowed because of 
the strict theoretical upper bound on the Higgs boson 
mass \cite{hmass}, and the decay receives dominant contributions 
from the SUSY QCD loops of 
flavor-changing interactions \cite{tch_mssm}. If the gluino
and squarks involved in the contributing SUSY QCD loops
are both light of order 100 GeV, the branching fraction 
could be enhanced to a level of $10^{-5}$. The branching
fraction falls off quickly for heavier sparticles in the
loops.

In this Letter we examine the R-parity-violating contributions to $t\to ch$. 
It is well-known that the R-parity conservation in SUSY theory, 
which implies the separate conservation of baryon number and 
lepton number, is put in by hand. R-parity violation ($\Rv$) 
can be made perfectly consistent with other fundamental principles 
such as gauge invariance, 
supersymmetry and renormalizability \cite{rv_mssm}. 
If R-parity violation is included
in the MSSM, $t\to ch$ will receive new contributions from the loops 
of R-parity-violating interactions. Such contributions can be 
significant for the following reasons. First, such contributing loops
only involve a single sparticle, {\it i.~e.}, a squark or a slepton.
The mass suppression in the loops thus becomes less severe than
that in the MSSM. 
Second, the relevant $\Rv$ couplings inducing $t\to ch$ involve
the third-generation fermions and are subject to rather weak bounds
from low-energy experiments.  We find that the branching fraction 
for $t\to c h$ from $\Rv$ contributions 
can be indeed as high as about $10^{-5}$ in some part 
of the parameter space, reaching a level of potentially
accessible at the LHC with a high luminosity.

We start our study by writing down the $\Rv$ superpotential of the MSSM 
\begin{equation} \label{poten} 
{\cal W}_{\Rv}=\frac{1}{2}\lambda_{ijk}L_iL_jE_k^c
+\lambda_{ijk}'L_iQ_jD_k^c
+\frac{1}{2}\lambda^{\prime\prime}_{ijk}U_{i}^cD_{j}^cD_{k}^c
+\mu_iL_iH_2,
\end{equation}
where $L_i(Q_i)$ and $E_i(U_i,D_i)$ are the left-handed lepton (quark) 
doublet and right-handed lepton (quark) singlet chiral superfields. 
$i,j,k$ are generation indices and $c$ denotes charge  conjugation. 
Note that $SU(2)_L$
and $SU(3)_C$ indices have been suppressed.
$H_{1,2}$ are the Higgs-doublets chiral superfields. 
The  $\lambda_{ijk}$ and $\lambda'_{ijk}$ are the lepton-number-violating 
($\Lv$) couplings, and 
$\lambda^{\prime\prime}_{ijk}$ the baryon-number-violating 
($\Bv$) couplings. Constraints on these couplings 
have been obtained from various low-energy 
processes \cite{barger,D-decay,apv,nu-mass,Z-decay,others,review} 
and their phenomenology at hadron and lepton colliders have been 
intensively investigated recently \cite{review,rv_coll}. 
Note that 
although it is theoretically possible to have both $\Bv$ and $\Lv$ 
interactions, the non-observation of proton decay prohibits their simultaneous
presence, at least in the first two fermion generations. 
We therefore assume the existence of either $\Lv$ couplings 
or $\Bv$ couplings, and investigate  their separate effects in 
top-quark decay $t\to ch$. 

We focus our attention only on the tri-linear supersymmetric $\Rv$ interactions
in Eq.~(\ref{poten}) and assume that the bi-linear terms $\mu_iL_iH_2$
can be rotated away by a field redefinition \cite{rv_mssm}. 
In this case the FCNC decay $t\to ch$
is induced by only the tri-linear $\Rv$ interactions via loops. 
Note that, in principle, there are also possible  
$\Rv$ terms in the soft-breaking part \cite{soft_rv}. In that case, 
it is no longer possible in general to rotate 
away the bi-linear terms \cite{soft_rv,lynne} and such
bi-linear terms will cause the mixing between the neutral Higgs 
bosons and the sneutrinos ($\tilde \nu$). 
As studied in \cite{tcsnu}, the FCNC decay $t\to c \tilde \nu$
can be induced by the one-loop diagrams of the tri-linear couplings. 
Followed by the oscillation $\tilde \nu\to h$ induced by the bi-linear 
$\Rv$ terms, one can also have $t\to c \tilde \nu \to c h$, which, however,
is only appreciable when $h$ and $\tilde \nu$ are nearly degenerate.
We will not consider this possibility further.

\begin{figure}[tb]
\vspace*{-6cm}
\centerline{
\psfig{figure=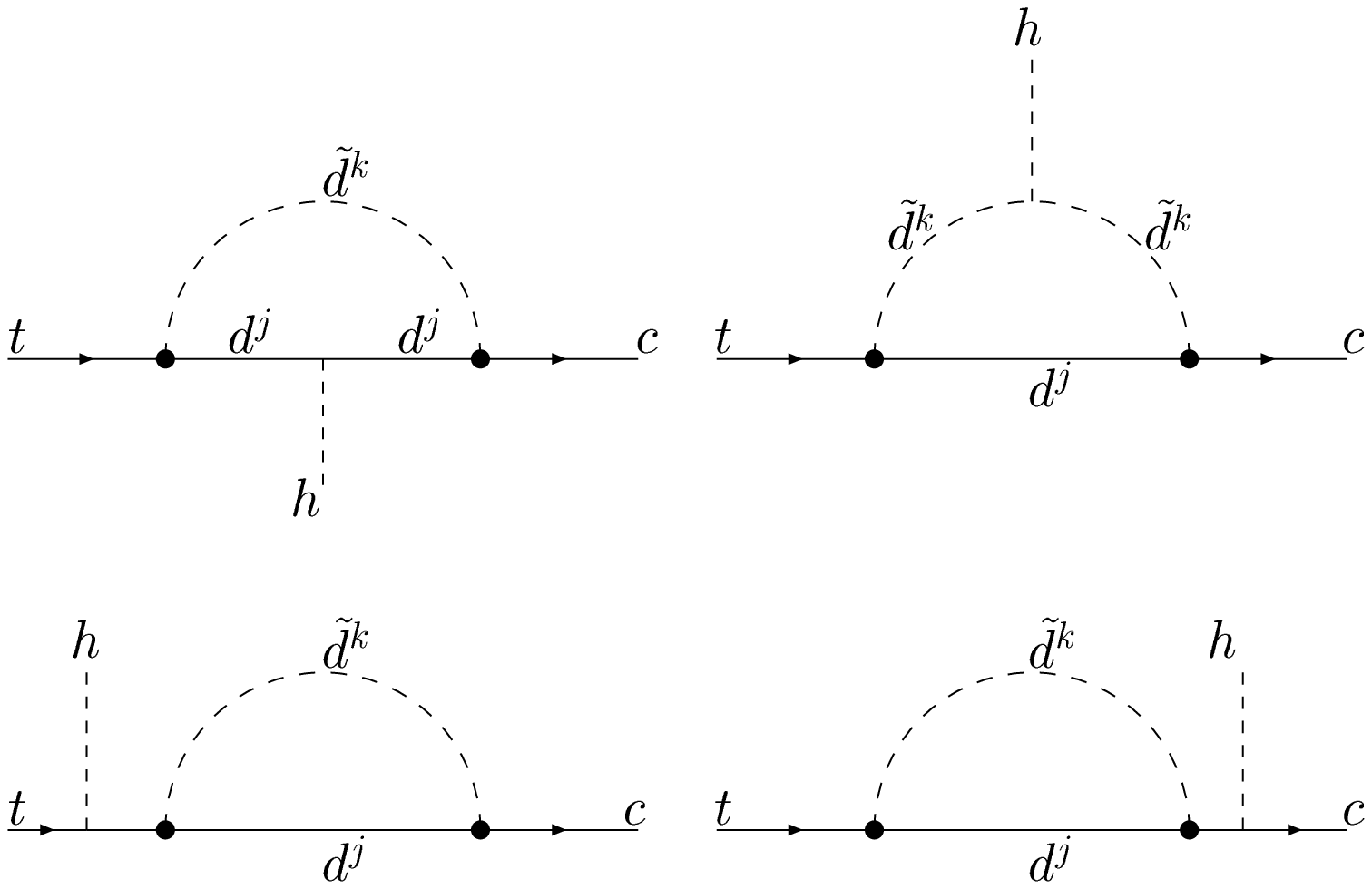,width=400pt,angle=0} }
\vspace*{-6.5cm}
\caption[]{Feynman diagrams of $t \to ch$ induced by  $\Bv$
  SUSY interactions.}
\label{fig1}
\end{figure}

In terms of the four-component Dirac notation, the Lagrangian of the
$\Lv$ couplings $\lambda'$ and $\Bv$ couplings $\lambda^{\prime\prime}$
are given by
\begin{eqnarray}
{\cal L}_{\lambda'}&=&-\lambda'_{ijk}
\left [\tilde \nu^i_L\bar d^k_R d^j_L+\tilde d^j_L\bar d^k_R\nu^i_L
       +(\tilde d^k_R)^*(\bar \nu^i_L)^c d^j_L\right.\nonumber\\
& &\hspace{1cm} \left. -\tilde e^i_L\bar d^k_R u^j_L
       -\tilde u^j_L\bar d^k_R e^i_L
       -(\tilde d^k_R)^*(\bar e^i_L)^c u^j_L\right ]+h.c.,\\
{\cal L}_{\lambda^{\prime\prime}}&=&-\frac{1}{2}\lambda^{\prime\prime}_{ijk}
\left [\tilde d^k_R(\bar u^i_R)^c d^j_R+\tilde d^j_R(\bar d^k_R)^c u^i_R
       +\tilde u^i_R(\bar d^j_R)^c d^k_R\right ]+h.c.
\end{eqnarray}
With $\Bv$ couplings, the decay $t\to ch$ can proceed 
through the loop diagrams shown in Fig.~1, where a down-type quark and 
a squark are involved in the loops. If the lepton number is violated instead, 
the decay $t\to ch$ can occur through the diagrams in Fig.~2,
where a down-type quark and a slepton (or a down-type squark and a
charged lepton) are in the loops.  We note that 
in the former case the products of  $\Bv$ couplings 
$\lambda^{\prime\prime}_{2jk}\lambda^{\prime\prime}_{3jk}$ are 
involved while in the latter case
the products of  $\Lv$ couplings $\lambda'_{i2k}\lambda'_{i3k}$ are involved. 

\begin{figure}[tb]
\vspace*{-6cm}
\begin{center}
\psfig{figure=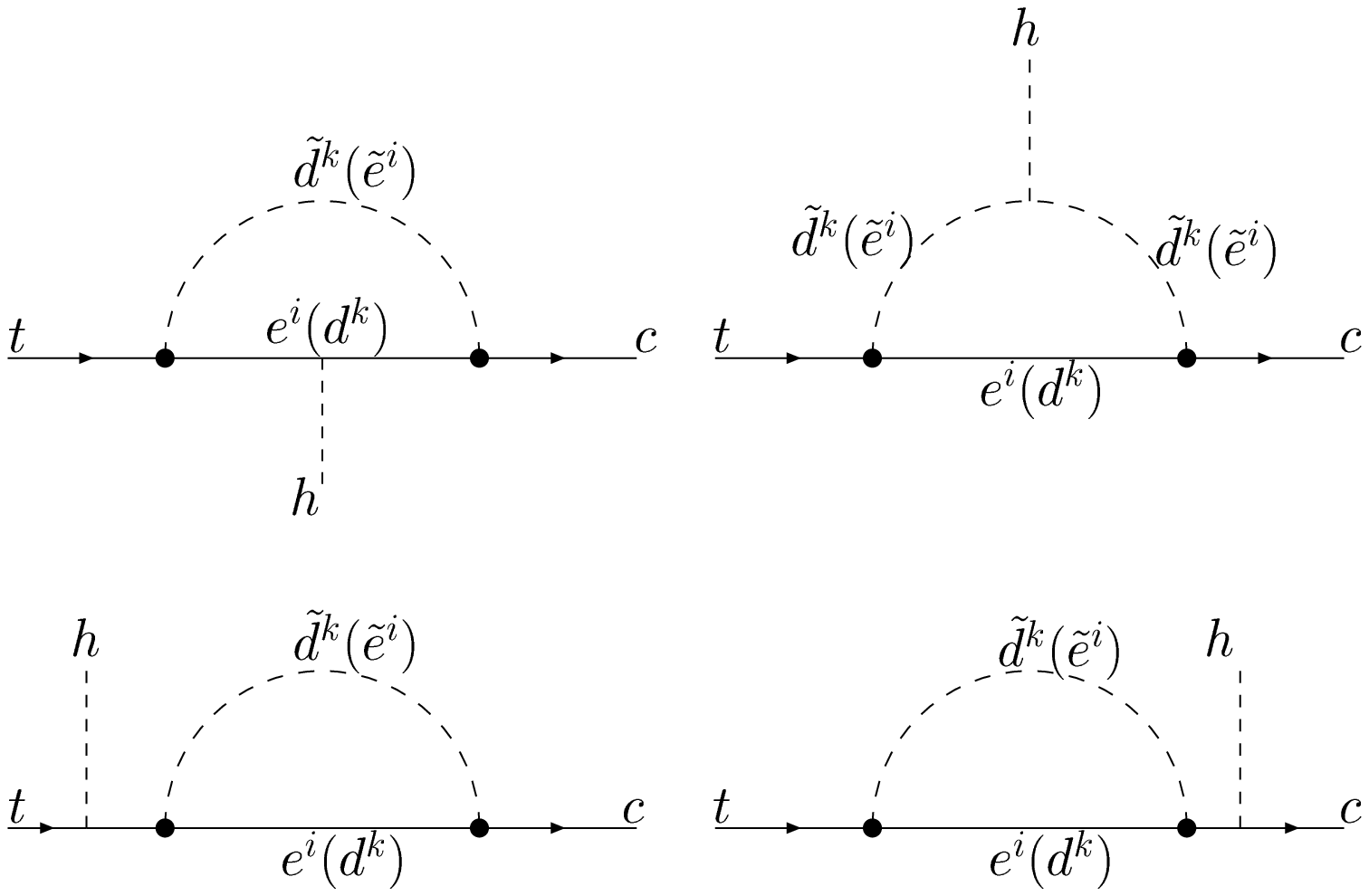,width=400pt,angle=0}
\end{center}
\vspace*{-7cm}
\caption{ Feynman diagrams of $t \to ch$ induced by  
  $\Lv$ SUSY interactions.}
\label{fig2}
\end{figure}
\vspace*{.5cm}

The induced effective $tch$ vertex is given by, in the cases of
$\Bv$ and $\Lv$ respectively,
\begin{eqnarray} \label{vertex}
V_{\rm tch}^{\not \! B}&=&
 ie\frac{\lambda^{\prime\prime}_{2jk}\lambda^{\prime\prime}_{3jk}}
{16\pi^2}\xi_c
\left ( F_L^{\not \! B} P_L+F_R^{\not \! B} P_R \right ),\\
V_{\rm tch}^{\not \! L} &=&
 ie\frac{\lambda'_{i2k}\lambda'_{i3k}}{16\pi^2}
\left (F_L^{\not \! L} P_L+F_R^{\not \! L} P_R \right ),
\end{eqnarray}
where $\xi_c=(N_c-1)!=2$ is the color factor and 
$P_{R,L}=\frac{1}{2}(1\pm \gamma_5)$. The factors $F_{L,R}^{\not \! B}$ 
from the contributions of Fig.~1 are given by
\begin{eqnarray}
F_L^{\not \! B}&=&Y_{d^j}m_t m_{d^j} (C_0+2C_{11})
 (-p_t,p_c,m_{d^j},m_{\tilde d^k},m_{d^j})\nonumber \\
& &+Y_{\tilde d^k_R}m_t (C_{11}-C_{12})
 (-p_t,k,m_{d^j},m_{\tilde d^k},m_{\tilde d^k})\nonumber \\
& &-Y_c \frac{m_t m_c}{m_t^2-m_c^2} B_1(p_t,m_{d^j},m_{\tilde d^k})
-Y_t \frac{ m_c^2}{m_c^2-m_t^2} B_1(p_c,m_{d^j},m_{\tilde d^k}),\\
F_R^{\not \! B}&=&-Y_{d^j}m_c m_{d^j} (C_0+2C_{12})
 (-p_t,p_c,m_{d^j},m_{\tilde d^k},m_{d^j})\nonumber \\
& &+Y_{\tilde d^k_R}m_c C_{12}
(-p_t,k,m_{d^j},m_{\tilde d^k},m_{\tilde d^k})\nonumber \\
& &-Y_c \frac{m_t^2 }{m_t^2-m_c^2} B_1(p_t,m_{d^j},m_{\tilde d^k})
-Y_t \frac{ m_t m_c}{m_c^2-m_t^2} B_1(p_c,m_{d^j},m_{\tilde d^k}).
\end{eqnarray}
The factors $F_{L,R}^{\not \! L}=F_{L,R}^{\not \! L}(\tilde d)
+F_{L,R}^{\not \! L}(\tilde e)$, where $F_{L,R}^{\not \! L}(\tilde d)$ 
and $F_{L,R}^{\not \! L}(\tilde e)$ are respectively the contributions of  
Fig.~2 by exchanging squarks  $\tilde d^k$ and sleptons  $\tilde e^i$,
are obtained by the substitutions
\begin{eqnarray}
F_{L,R}^{\not \! L}(\tilde d)&=&
F_{R,L}^{\not \! B}\left \vert_{d^j\to e^i}, \right.\\
F_{L,R}^{\not \! L}(\tilde e)&=&
F_{R,L}^{\not \! B}
\left \vert_{d^j\to d^k, \tilde d^k_R\to \tilde e^i_L, m_{d^j}\to -m_{d^k}}.
 \right.
\end{eqnarray}
In the above expressions the sum over family indices $i,j, k=1,2,3$ is 
implied. $p_t$, $p_c$ and $k$ are the momenta of the top quark, charm quark
and Higgs boson, respectively. The functions $B_1$ and $C_{ij}$ are the
conventional 2- and 3-point Feynman integrals \cite{bcs}, and their 
functional arguments are 
indicated in the bracket following them.  The constants like $Y_{d^j}$
are the Yukawa couplings of the corresponding particles given by
\begin{eqnarray}
Y_{d}&=&\frac{m_{d}\sin\alpha}{2m_W \sin\theta_W \cos\beta},\\
Y_{u}&=&\frac{m_{u}\cos\alpha}{2m_W \sin\theta_W \sin\beta},\\
Y_{\tilde d_R}&=&-\frac{1}{3}m_Z \tan\theta_W \sin(\alpha+\beta)
                +\frac{m_d^2}{m_W\sin\theta_W} \frac{\sin\alpha}{\cos\beta},\\
Y_{\tilde e_L}&=&-\frac{m_Z}{\cos\theta_W \sin\theta_W}
                 (\frac{1}{2}-\sin^2\theta_W) \sin(\alpha+\beta)
                +\frac{m_e^2}{m_W\sin\theta_W} \frac{\sin\alpha}{\cos\beta},
\end{eqnarray}
where $\theta_W$ is the weak mixing angle, $\alpha$ the neutral Higgs
boson mixing angle, and $\tan\beta=v_2/v_1$ the ratio of the two vacuum
expectation values in the Higgs sector \cite{mssm}.   

The ultraviolet divergences are contained in the Feynman integral
$B_1$. It is easy to check that all the ultraviolet divergences  
cancel as a result of renormalizability of MSSM.

\begin{table}
\caption{ Current $2\sigma$ bounds on $\lambda^{\prime\prime}_{2jk}$, 
        $\lambda^{\prime\prime}_{3jk}$, $\lambda'_{i2k}$ and $\lambda'_{i3k}$.}
\vspace{0.1in}
\begin{tabular}{lll}
  couplings & bounds & sources \\ \hline
$\lambda'_{121}, \lambda'_{122}, \lambda'_{123}$ & 
  $0.043\times ({m_{\tilde d_R}}/{100\ {\rm GeV}})$& 
  charged current universality \cite{barger} \\
$\lambda'_{221}$ & $0.18\times ({m_{\tilde d_R}}/{100\ {\rm GeV}})$&
$\nu_{\mu}$ deep inelastic scattering \cite{barger,D-decay}\\
$\lambda'_{222},\lambda'_{223}$ & 
$0.21\times ({m_{\tilde d_R}}/{100\ {\rm GeV}})$&
$D$-meson decay \cite{D-decay} \\
$\lambda'_{321}, \lambda'_{322},\lambda'_{323}$ &
$0.52\times ({m_{\tilde d_R}}/{100\ {\rm GeV}})$& 
  $D_s$ decay \cite{D-decay},\\
$\lambda'_{131}$ & $0.019\times ({m_{\tilde t_L}}/{100\ {\rm GeV}})$&
atomic parity violation \cite{barger,apv}\\
$\lambda'_{132}$ &$ 0.28\times ({m_{\tilde t_L}}/{100\ {\rm GeV}})$&
asymmetry in $e^+e^-$ collision \cite{barger}\\
$\lambda'_{133}$ & $0.0014\times \sqrt {{m_{\tilde b}}/{100\ {\rm GeV}}}$
                   ~~~~& neutrino mass \cite{nu-mass}\\
$\lambda'_{231}$ & $0.18\times ({m_{\tilde b_L}}/{100\ {\rm GeV}})$ &
          $\nu_{\mu}$ deep inelastic scattering  \cite{barger,D-decay}\\
                 $\lambda'_{232}$& $0.56$ & $Z$ decays \cite{Z-decay} \\
$\lambda'_{233}$ & $0.15\times \sqrt {{m_{\tilde b}}/{100\ {\rm GeV}}}$&
                 neutrino mass \cite{nu-mass} \\ 
$\lambda'_{331},\lambda'_{332},\lambda'_{333}$ & $0.45$ & 
                   $Z$ decays \cite{Z-decay} \\
$\lambda^{\prime\prime}_{212},\lambda^{\prime\prime}_{213},
\lambda^{\prime\prime}_{223}$&$1.23$ & 
                 perturbativity \cite{pertur} \\ 
$\lambda^{\prime\prime}_{312}, \lambda^{\prime\prime}_{313}, 
\lambda^{\prime\prime}_{323}$~~~~
     & $1.0$&  $Z$ decays \cite{Z-decay}
\end{tabular}
\end{table}

As a good approximation of negelecting all the fermion masses 
but $m_t$, the expression can be substantially simplified.
The only contributing diagram in this approximation is that 
of the Higgs boson coupled to the sfermion. The partial decay 
width is then given by a simple form
\begin{equation} \label{width} 
\Gamma(t\to ch)= {K^2(m_t^2-m_h^2)^2\over 32\pi m_t}
\left|{(C_{11}-C_{12})
(-p_t,k,0,m_{\tilde f},m_{\tilde f})\over 16\pi^2}\right|^2,
\end{equation}
where 
\[ K= \left\{ \begin{array}{ll}
e\ \xi_c\ Y_{\tilde d^k_R}
\lambda^{\prime\prime}_{2jk}\lambda^{\prime\prime}_{3jk} 
& \mbox{for ${\not \! B}$}\\ 
e\ Y_{\tilde e^i_L}
\lambda^{\prime}_{i2k}\lambda^{\prime}_{i3k} 
& \mbox{for ${\not \! L}$.}
\end{array}
\right. \]
This expression is accurate at a level of a few percent.

Now we present the numerical results for $Br(t\to ch)$
with the full expression.
For the SM parameters involved, we take 
$m_t=175$ GeV, $m_Z=91.187$ GeV, $m_W=80.3$ GeV, $\alpha=1/128$,
$\sin^2\theta_W=0.232$, $m_c=1.7$ GeV, 
$m_b=4.7 $ GeV, and $m_s=0.17 $ GeV.
The results are not sensitive to the light fermion masses.
%$m_e=0.5$ MeV, $m_{\mu}=105.7$ MeV and $m_{\tau}=1.7$ GeV.
The SUSY parameters involved are the following:
\begin{itemize}
\item[{\rm(1)}]  
$\Bv$ couplings $\lambda^{\prime\prime}_{2jk}$ and 
$\lambda^{\prime\prime}_{3jk}$, 
and $\Lv$ couplings  $\lambda'_{i2k}$ and $\lambda'_{i3k}$:
The current bounds for all $\Rv$ couplings are summarized in 
Ref.~\cite{review}. In Table 1 we only list those relevant to our study.  
We note that the current bounds for the $\Bv$ couplings are generally
quite weak. 

\item[{\rm(2)}] Slepton and down-type squark masses: The
off-diagonal terms in the sfermion-mass matrices
are proportional to the mass of the corresponding fermion, 
and they are thus relatively small for the down-type squarks 
and sleptons. Only for large $\tan\beta$, the mass splitting
of the sbottoms (staus) becomes sizeable. Nevertheless, the
inclusion of the mass splittings for sbottom (stau)
does not affect our results appreciably.  
In our numerical illustration, we thus assume a 
common mass for all squarks (sleptons).

\item[{\rm(3)}] Higgs masses and mixing angles of the Higgs sector:
At tree level the Higgs sector of the MSSM is determined by two free 
parameters, {\it e.~g.}, the CP-odd Higgs boson mass $m_A$ and 
$\tan\beta$. 
When radiative corrections are included, several other parameters 
enter through the loops \cite{hmass}. 
In phenomenological analysis it is usually to assume
a common squark mass scale $m_{\tilde q}$ and thus the dominant one-loop
corrections can be parameterized by a single parameter $\epsilon$
\cite{epsilon}, 
\begin{eqnarray}
\epsilon=\frac{3g^2}{8\pi^2 m_W^2}m_t^4 
         \ln \left ( 1+\frac{m_{\tilde q}^2}{m_t^2} \right ).
\end{eqnarray}
Then the Higgs masses and mixing angle $\alpha$ are obtained by
\begin{eqnarray}
m^2_{H,h}&=&\frac{1}{2}\left [m_A^2+m_Z^2+\epsilon/\sin^2\beta\right ] 
                                                           \nonumber\\
         & & \pm \frac{1}{2}\left \{ 
\left [ \left (m_A^2-m_Z^2 \right )\cos 2\beta+\epsilon/\sin^2\beta \right ]^2
       +\left (m_A^2+m_Z^2 \right )^2 \sin^2 2\beta \right \}^{1/2}, \\
\tan 2\alpha &=& \frac{\left (m_A^2+m_Z^2 \right ) \sin 2\beta }
            { \left (m_A^2-m_Z^2 \right )\cos 2\beta+\epsilon/\sin^2\beta}.
\end{eqnarray}
In our calculation we adopt the standard convention \cite{mssm,epsilon}, 
in which $-\pi/2\le\alpha\le0$ and $0\le\beta\le \pi/2$ so that as
$m_A\to \infty$, we have $\alpha \to \beta-\pi/2$ and thus the $h$ couplings 
approach to those of the SM Higgs boson while other Higgs bosons 
($A, H, H^{\pm}$) decouple.
  
The current constraints on the parameter space of the Higgs sector were
given by the LEP experiments \cite{LEP_higgs}. The region of small
 $\tan\beta$ and low Higgs mass $m_A$ has been excluded, depending
on the extent of top-squark mixing. In any case the robust bounds 
 $m_h>83.4$ GeV and $m_A>83.8$ GeV have been obtained 
at 95\% confidence level for $\tan\beta$ greater than 0.8.
\end{itemize}
 
The decay rate increases quadratically with the relevant $\lambda'$ or 
$\lambda^{\prime\prime}$  couplings and decreases with the 
increase of the sparticle mass.
Using the upper limits of the relevant $\Rv$ couplings in Table 1, 
we present the maximum values of the branching fraction in Figs.~$3-5$. 

%%%%%%%%%%%%%%%%%%%%%%%%%
\begin{figure}[tb]
\vspace*{-2cm}
\begin{center}
\psfig{figure=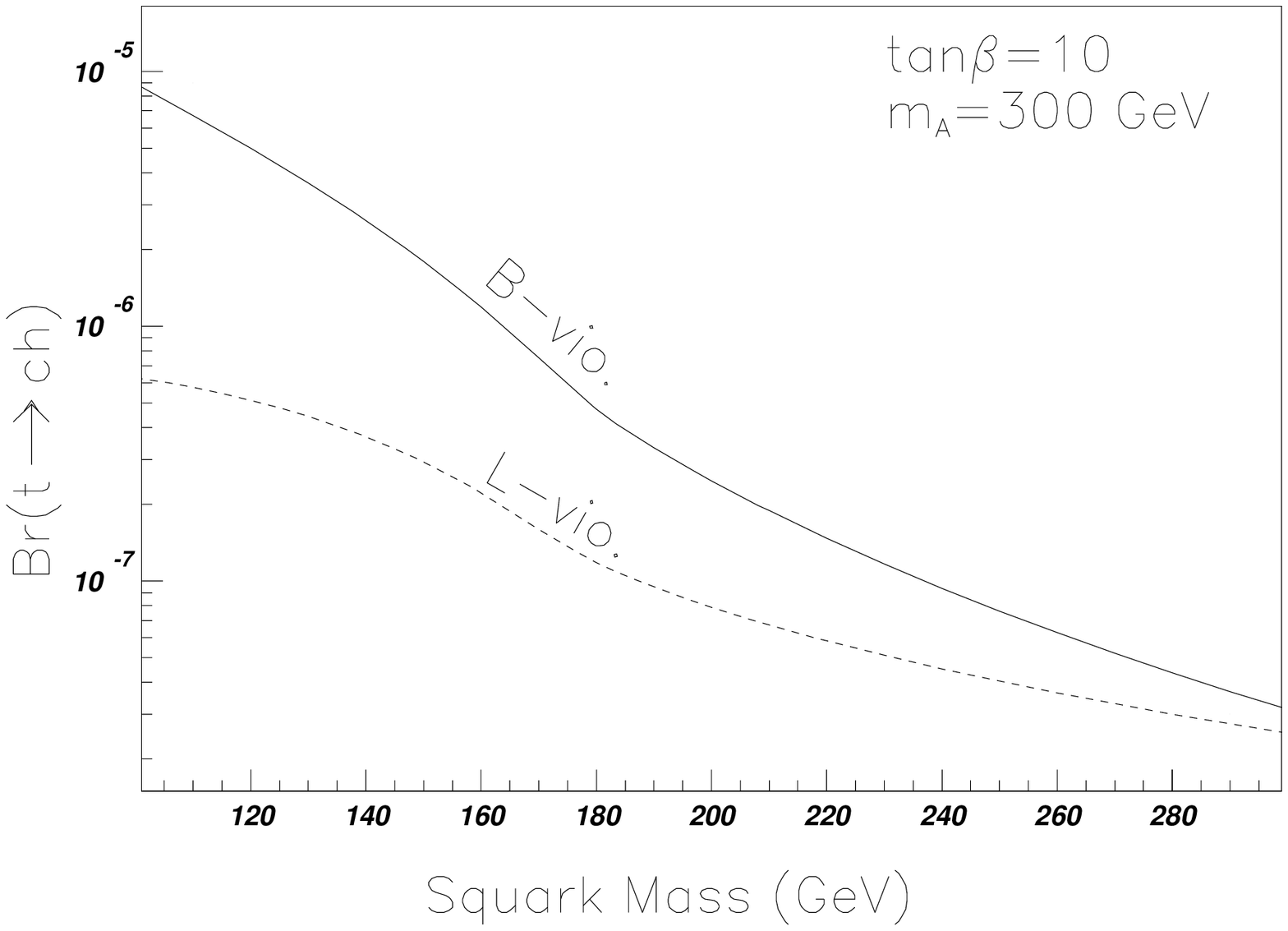,width=350pt,angle=-90}
\end{center}
\vspace*{-2cm}
\caption{ $Br(t \to ch)$ versus squark mass
       for  $m_A=300$ GeV and  $\tan\beta=10$.
The solid and dashed curves correspond to 
B-violation and L-violation, respectively.}
\label{fig3}
\end{figure}

Fig.~3 shows the dependence of $Br(t \to ch)$ on squark mass for 
$m_A=300$ GeV and  $\tan\beta=10$. The solid and dashed curves 
correspond to B-violation and L-violation, respectively. 
With the parameter values in this figure, the value of $m_h$ is
within the range $93 \sim 106$ GeV corresponding to the squark 
mass range $100\sim 300$ GeV. The results manifest the 
decoupling property of the MSSM, {\it i.~e.}, 
the contributions drop with the increase of the squark mass. 
We found that when squarks become sufficiently heavy ($\gsim 500$ GeV),
they decouple quickly and the branching fraction
goes down like $1/m_{\tilde q}^4$ asymptotically.
From Fig.~3 one sees that in the $\Bv$ case, for a squark as light as 100 GeV, 
the branching fraction can reach the level of $10^{-5}$. In the $\Lv$ case, 
however, the contributions are below the level of $10^{-6}$ because of the
current stringent bounds on the relevant couplings in Table 1. 
Thus in the following we do not present the $\Lv$ contributions
any more.   

%%%%%%%%%%%%%%%%%%%%%%%%%
\begin{figure}[tb]
\vspace*{-2cm}
\begin{center}
\psfig{figure=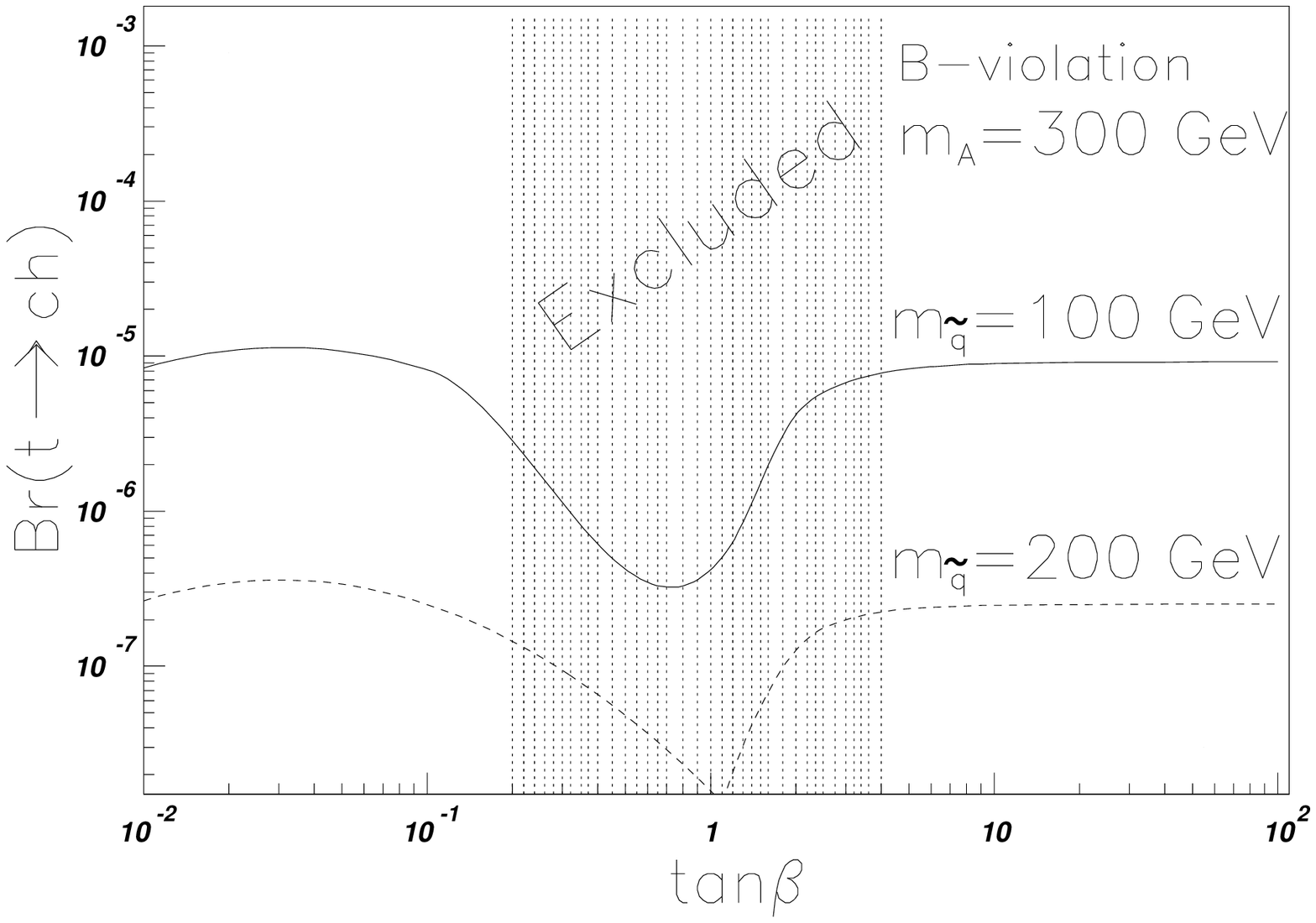,width=350pt,angle=-90}
\end{center}
\vspace*{-2cm}
\caption{ $\Bv$ contributions to $Br(t \to ch)$ 
        versus $\tan\beta$ for $m_A=300$ GeV.
The solid and dashed curves correspond to 
squark mass of 100 GeV and 200 GeV, respectively.
The hatched region is excluded by LEP experiments.}
\label{fig4}
\end{figure}

Fig.~4 shows the $\Bv$ contributions to $Br(t \to ch)$ as a function
of $\tan\beta$ for $m_A=300$ GeV. The solid and dashed curves 
correspond to squark mass of 100 GeV and 200 GeV, respectively.
The hatched region is excluded by LEP experiments \cite{LEP_higgs}. 
Note that in the case of maximal top-squark mixing, the very small  
$\tan\beta$ region ($\lsim 0.2$) is still allowed by LEP experiments
and in this region the branching fraction can also be significant. 
Regarding $m_h$, in the small 
$\tan\beta$ range $0.01\lsim \tan\beta\lsim 0.2$, 
we have $91\lsim m_h \lsim 92$ GeV ($91\lsim m_h \lsim 93$) 
for squark mass of 100 GeV (200 GeV).
In the large $\tan\beta$ range $4\lsim \tan\beta\lsim 100$, 
we have $84 \lsim m_h \lsim 94$ GeV ($91 \lsim m_h \lsim 101$ GeV)
for squark mass of 100 GeV (200 GeV), correspondingly.

%%%%%%%%%%%%%%%%%%%%%%%%%
\begin{figure}[tb]
\vspace*{-2cm}
\begin{center}
\psfig{figure=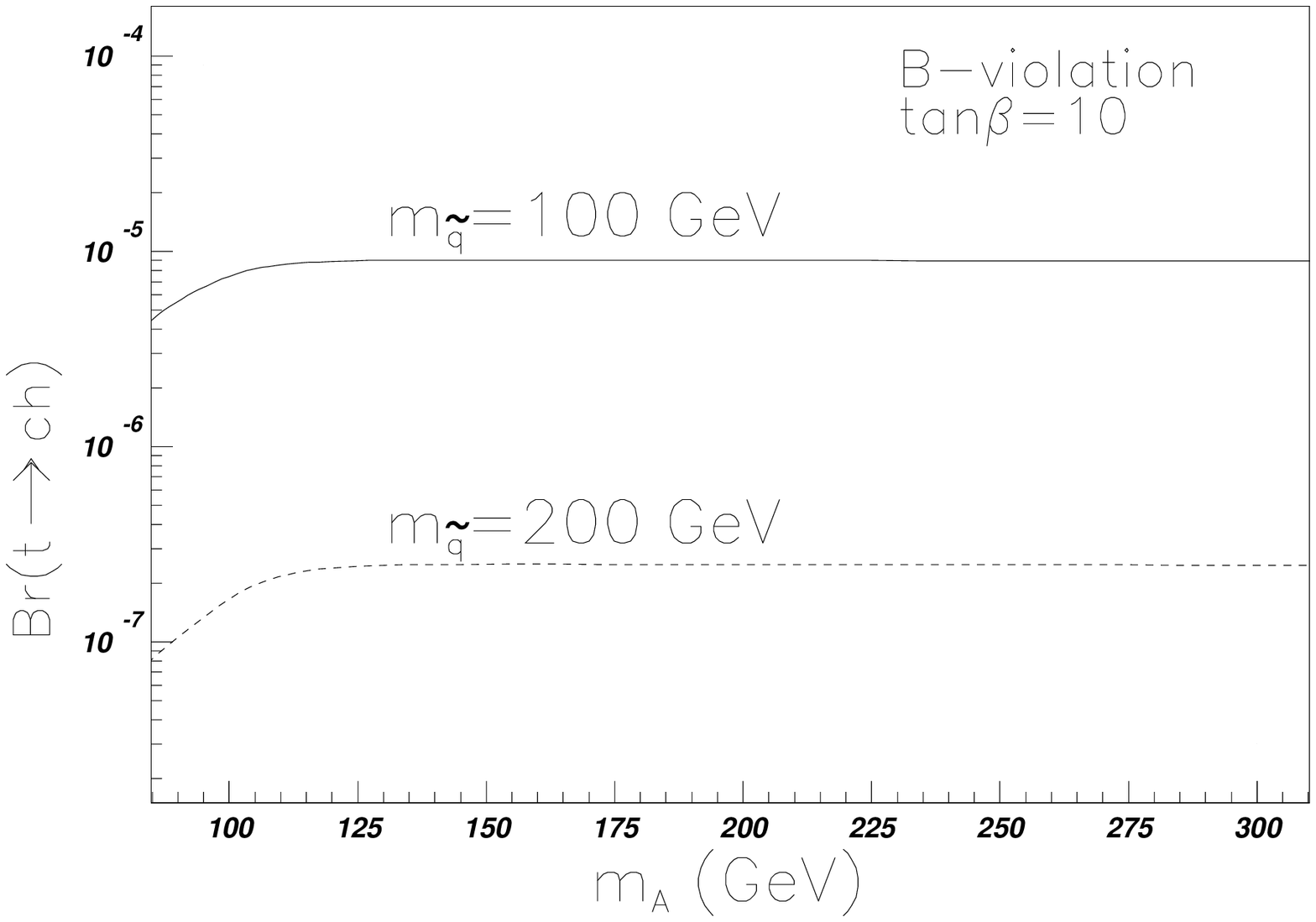,width=350pt,angle=-90}
\end{center}
\vspace*{-2cm}
\caption{  $\Bv$ contributions to $Br(t \to ch)$ 
        versus $m_A$ for  $\tan\beta=10$.
The solid and dashed curves correspond to 
squark mass of 100 GeV and 200 GeV, respectively.}
\label{fig5}
\end{figure}
    
Fig.~5  is the $\Bv$ contribution to $Br(t \to ch)$ versus $m_A$ for 
$\tan\beta=10$. The solid and dashed curves correspond to 
squark mass of 100 GeV and 200 GeV, respectively.
The range of $m_h$ is $83 \lsim m_h \lsim 93$ GeV and
$85 \lsim m_h \lsim 100$ GeV for squark mass of 100 GeV, 
200 GeV, respectively.
This figure shows that the dependence of the branching
fraction on $m_A$ is quite mild.

A couple of remarks are in order regarding the above results:
\begin{itemize}
\item[{\rm(1)}] The branching fraction is very sensitive to the relevant
$\Rv$ couplings, {\it e.~g.}, in the $\Bv$ case, is proportional to
$\left (\lambda^{\prime\prime}_{2jk} \lambda^{\prime\prime}_{3jk}\right )^2$. 
The relaxation of some bounds in Table 1 will result in large enhancement
of the  branching fraction. For example, so far no experimental bound for 
$\lambda^{\prime\prime}_{2jk}$ has been reported and we used the theoretical
perturbativity bound ($\simeq 1.23$) \cite{pertur}. Such a bound is derived 
from the assumption that the gauge groups unify at $M_U=2\times 10^{16}$ GeV 
and the Yukawa couplings $Y_t, Y_b$ and $Y_{\tau}$ remain in the perturbative 
domain in the whole range up to $M_U$, {\it i.~e.}, $Y_i(\mu)<1$ for 
$\mu<2\times 10^{16}$ GeV.  But there is no a priori reason to take this 
theoretical assumption. So the bound on $\lambda^{\prime\prime}_{2jk}$ 
could be even weaker. 
\item[{\rm(2)}] The decay rate is proportional to a product of
two $\Rv$ couplings, and thus the non-zero decay requires the  
coexistence of at least two couplings. 
Although we do not expect any stringent constraints on the
coupling products involving the third-generation fermions, 
it would be interesting to seek for other possible FCNC processes at
low energies which may be induced by such coexistence of 
more than one $\Rv$ coupling.   
\item[{\rm(3)}]  
We assumed a common mass for all squarks or sleptons for illustration
of the numerical results. Our results are not very 
sensitive to the mass differences of the sfermions
because the involved $\Rv$ couplings,
$\lambda^{\prime\prime}$ or $\lambda^{\prime}$, are not unitary 
in general so that the delicate cancellations (like the GIM mechanism)
do not generally occur.
\item[{\rm(4)}] Although in our analysis we varied $\tan\beta$ freely and
only considered the direct LEP constraints, we note that the extreme case 
of a very small or very large  $\tan\beta$ value may be severely disfavored 
by some theoretical arguments and/or by some indirect experimental 
constraints like those from $b\to s\gamma$ \cite{bsg}.
Since the branching ratio of $t\to ch$ could be significantly 
enhanced also in 
the intermediate range of $\tan\beta$ (say $4\lsim \tan\beta\lsim 30$),
our conclusion will not be affected by such indirect experimental 
constraints.  
\item[{\rm(5)}] Finally, we point out that we only presented the analysis 
for the most interesting decay channel $t\to ch$ because kinematically 
it is always allowed. Other decay channels, $t\to cH$ or $t\to cA$, are 
quite likely to be suppressed by phase space or even kinematically forbidden.
If the channel $t\to c A$ is kinematically allowed, the branching ratio 
$Br(t\to c A)$ can also be significant for a very large $\tan\beta$ 
$(\gsim 40)$ because its coupling to bottom quark is proportional to 
$\frac{m_b}{m_W} \tan\beta$. Since its coupling to top quark is proportional 
to $\frac{m_t}{m_W} \cot\beta$, $Br(t\to c A)$ could be enhanced significantly
for a small $\tan\beta$. However, for such a light Higgs boson $A$ 
$(m_A \lsim 175~{\rm GeV})$, 
the small $\tan\beta$ region has been almost completely 
excluded by LEP experiments\cite{LEP_higgs}.
\end{itemize}

\vspace{.5cm}
In summary, we found that the FCNC decay $t \to ch$ can be significantly 
enhanced relative to that of the SM in SUSY theories with R-parity
violation. The branching fraction depends quadratically upon the
products of $\Rv$ couplings, and scales with the heavy mass
of the sfermions in the loops as $m^{-4}_{\tilde f}$, and it
can be at least as large as that in the MSSM.
In the optimistic scenario that the involved couplings take their 
current upper bounds and $m_{\tilde f}\approx 100$ GeV, 
the branching fraction can be as large as $10^{-5}$, which is potentially 
accessible at the LHC with a high luminosity.

\acknowledgments
G.E. would like to thank S. Bar-Shalom for very helpful discussions.
G.E. is supported in part by the Israel Science Foundation
and by the U.S-Israel Binational Science Foundation.
The work of T.H. is supported in part 
by a DOE grant No. DE-FG02-95ER40896 and by Wisconsin Alumni Research
Foundation.  
X.Z. is supported by National Natural Science Foundation of China.


\vspace{.5cm}

\begin{thebibliography}{99}
\bibitem{tcvh_sm} G.~Eilam, J.~L.~Hewett and A.~Soni, \PRD44, 1473 (1991),
erratum: \PRD59, 039901 (1999).
 %%CtITATION = PHRVA,D44,1473;%%
\bibitem{tch_sm} B.~Mele, S.~Petrarca, A.~Soddu, \PLB435, 401 (1998);
erratum in~\cite{tcvh_sm}.
 %%CITATION = PHLTA,B435,401;%%
\bibitem{tcv_mc} 
  T.~Han, R.~D.~Peccei, and X.~Zhang, \NPB454, 527 (1995);
 %%CITATION = NUPHA,B454,527;%%  
  T.~Han, K.~Whisnant, B.-L.~Young, and X.~Zhang, \PRD55, 7241 (1997); 
                                                  \PLB385, 311 (1996); 
 %%CITATION = PHRVA,D55,7241;%%
 %%CITATION = PHLTA,B385,311;%%
  M.~Hosch, K.~Whisnant, and  B.-L.~Young, \PRD56, 5725 (1997).
 %%CITATION = PHRVA,D56,5725;%%
\bibitem{tcvmore} 
E. Malkawi and T. Tait, \PRD54, 5758 (1996);
 %%CITATION = PHRVA,D54,5758;%%
T. Tait and C.P. Yuan,  \PRD55, 7300 (1997);
 %%CITATION = PHRVA,D55,7300;%%
T. Han, M. Hosch, K. Whisnant, B.-L. Young and X. Zhang, \PRD58, 073008 (1998);
 %%CITATION = PHRVA,D58,073008;%%
F. del Aguila and J.A. Aguilar-Saavedra, \PLB462, 310 (1999); 
 %%CITATION = PHLTA,B462,310;%%
F. del Aguila and J.A. Aguilar-Saavedra, \NPB576, 56 (2000).
 %%CITATION = NUPHA,B576,56;%% 
\bibitem{tch_mc} J.~A.~Aguilar-Saavedra and G.~C.~Branco, \PLB495, 347 (2000).
 %%CITATION = PHLTA,B495,347;%%
\bibitem{mIII} For top-quark FCNC in extended Higgs sectors, 
 see, {\it e.~g.}, T.P. Cheng and M. Sher, \PRD35, 3484 (1987);
 %%CITATION = PHRVA,D35,3484;%%
 B.Grz\c{a}dkowski, J.~F.~Gunion, and P.~Krawczyk,  \PLB268, 106 (1991);
 %%CITATION = PHLTA,B268,106;%%
 Ref.~\cite{tcvh_sm};
 A. Antaramian, L. Hall and A. Rasin, \PRL69, 1871 (1992); 
 %%CITATION = PRLTA,69,1871;%% 
 N.~G.~Deshpande, B.~Margolis, and H.~Trottier, \PRD45, 178(1992);
 %%CITATION = PHRVA,D45,178;%%
 W.-S. Hou, \PLB296, 179 (1992);
 %%CITATION = PHLTA,B296,179;%%
 M. Luke and M.J. Savage, \PLB307, 387 (1993);
 %%CITATION = PHLTA,B307,387;%%
 L. Hall and S. Weinberg, \PRD48, R979 (1993);
 %%CITATION = PHRVA,D48,R979;%%
 A. Atwood, L. Reina and A. Soni, \PRL75, 3800 (1995); 
         \PRD53, 1199 (1996); \PRD55, 3156 (1997);
 S.~B\'{e}jar, J.~Guasch and J.~S\`{o}la, hep-ph/0011091.
 %%CITATION = PRLTA,75,3800;%% 
 %%CITATION = PHRVA,D53,1199;%%
 %%CITATION = PHRVA,D55,3156;%%
\bibitem{tcv_mssm}  For $t \to cV$ in the MSSM, see,  {\it e.~g.},
   C.~S.~Li, R.~J.~Oakes and J.~M.~Yang, \PRD49, 293 (1994); 
 %%CITATION = PHRVA,D49,293;%%
   G.~Couture, C.~Hamzaoui and H.~Konig, \PRD52, 1713 (1995); 
 %%CITATION = PHRVA,D52,1713;%%
   J.~L.~Lopez, D.~V.~Nanopoulos and R.~Rangarajan, \PRD56, 3100  (1997);
 %%CITATION = PHRVA,D56,3100;%%
   G.~M.~de Divitiis, R.~Petronzio and L.~Silvestrini, \NPB504, 45 (1997).
 %%CITATION = NUPHA,B504,45;%%
\bibitem{tch_mssm}  For $t \to ch$ in the MSSM, see,  {\it e.~g.},
  J.~M.~Yang and C.~S.~Li, \PRD49, 3412 (1994); 
 %%CITATION = PHRVA,D49,3412;%%
  J.~Guasch, and J.~S\`{o}la, \NPB562, 3 (1999);
 %%CITATION = NUPHA,B562,3;%% 
S.~B\'{e}jar, J.~Guasch and J.~S\`{o}la, hep-ph/0101294.
\bibitem{tcv_rv} For $t \to cV$ in the $\Rv$ MSSM, see,  {\it e.~g.},
 J.~M.~Yang, B.-L.~Young and X.~Zhang, \PRD58, 055001 (1998).
 %%CITATION = PHRVA,D58,055001;%% 
\bibitem{tcsnu} S.~Bar-Shalom, G.~Eilam, and A.~Soni,  \PRD60, 035007 (1999).
 %%CITATION = PHRVA,D60,035007;%%
\bibitem{tcv_tc}  For $t \to cV$ in technicolor models, see,  {\it e.~g.}, 
 X.L.~Wang {\it et al.},  \PRD50, 
5781 (1994); \JPG20, L91 (1994); \CTP24, 359 (1995).
 %%CITATION = PHRVA,D50,5781;%%
 %%CITATION = JPHGB,20,L91;%%
 %%CITATION = CTPMD,24,359;%%
\bibitem{mssm} H.~E.~Haber and G.~L.~Kane, Phys. Rep. {\bf 117}, 75  (1985); 
%%CITATION = PRPLC,117,75;%%
               J.~F.~Gunion and H.~E.~Haber, \NPB272, 1  (1986).
%%CITATION = NUPHA,B272,1;%% 
\bibitem{hmass} M. Berger, \PRD41, 225 (1990);
 %%CITATION = PHRVA,D41,225;%%
                H.~E.~Haber and R.~Hempfling, \PRL66, 1815 (1991);
 %%CITATION = PRLTA,66,1815;%%
                J. Ellis, G. Ridolfi and F. Zwerner, \PLB257, 83 (1991);
 %%CITATION = PHLTA,B257,83;%%
                Y. Okada, M. Yamaguchi and T. Yanagida, \PLB262, 54 (1991). 
 %%CITATION = PHLTA,B262,54;%%
\bibitem{rv_mssm}
         L.~Hall and M.~Suzuki, \NPB231, 419 (1984); 
 %%CITATION = NUPHA,B231,419;%%
         J.~Ellis {\it et al.}, \PLB150, 142 (1985);
 %%CITATION = PHLTA,B150,142;%%
         G.~Ross and J.~Valle, \PLB151, 375 (1985);
 %%CITATION = PHLTA,B151,375;%%
         S.~Dawson, \NPB261, 297 (1985);
 %%CITATION = NUPHA,B261,297;%%
         R.~Barbieri and A.~Masiero, \NPB267, 679 (1986);
 %%CITATION = NUPHA,B267,679;%% 
        R.N. Mohapatra, \PRD34, 3457 (1986).
 %%CITATION = PHRVA,D34,3457;%%
\bibitem{barger}  V.~Barger, G.~F.~Giudice and T.~Han, \PRD40, 2978 (1989).
 %%CITATION = PHRVA,D40,2978;%%  
\bibitem{D-decay} G.~Bhattacharyya and D.~Choudhury, \MPLA10,  1699 (1995).
%%CITATION = MPLAE,A10,1699;%%
\bibitem{apv} S.~C.~Bennett, and C.~E.~Wieman, \PRL82, 2484 (1999).
 %%CITATION = PRLTA,82,2484;%% 
\bibitem{nu-mass} R.~M.~Godbole, P.~Roy and X.~Tata, \NPB401, 67 (1993).
 %%CITATION = NUPHA,B401,67;%% 
\bibitem{Z-decay} 
 J.~M.~Yang, hep-ph/9905486;
 %%CITATION = HEP-PH 9905486;%%
 O.~Lebedev, W.~Loinaz, and T.~Takeuchi, hep-ph/9910435; hep-ph/9911479; 
 %%CITATION = HEP-PH 9910435;%%
 %%CITATION = HEP-PH 9911479;%% 
 G.~Bhattacharyya, J.~Ellis and K.~Sridhar, \MPLA10,1583 (1995);
 %%CITATION = MPLAE,A10,1583;%%
 G.~Bhattacharyya, D.~Choudhury and K.~Sridhar, \PLB355, 193  (1995). 
 %%CITATION = PHLTA,B355,193;%%
\bibitem{others}  
     K.~Agashe and M.~Graesser, \PRD54, 4445 (1996);
 %%CITATION = PHRVA,D54,4445;%%
     K.~S.~Babu and R.~N.~Mohapatra, \PRL75, 2276 (1995);  
  %%CITATION = PRLTA,75,2276;%%
     J.~Jang, J.~K.~Kim and J.~S.~Lee, \PRD55, 7296 (1997);
 %%CITATION = PHRVA,D55,7296;%%
     G.~Bhattacharyya and A.~Raychaudhuri, \PRD57, 3837 (1998);
 %%CITATION = PHRVA,D57,3837;%%
     P.~Slavich, hep-ph/0008270.
 %%CITATION = HEP-PH 0008270;%% 
\bibitem{review} For a recent review on $\Rv$ phenomenology, 
    see, {\it e.~g.}, B.~Allanach {\it et al.}, 
   (edited by H.~Dreiner ),  hep-ph/9906224. 
 %%CITATION = HEP-PH 9906224;%% 
\bibitem{rv_coll}
         H.~Dreiner and G.G.~Ross,  \NPB365, 597 (1991);
 %%CITATION = NUPHA,B365,597;%%
         H.~Dreiner and R.J.N. Phillips, \NPB367, 591 (1991);
 %%CITATION = NUPHA,B367,591;%%
  J.~Erler, J.~L.~Feng and N.~Polonsky, \PRL78, 3063 (1997);
 %%CITATION = PRLTA,78,3063;%%  
  A.~Datta,  J.~M.~Yang, B.-L.~Young and X.~Zhang, \PRD56, 3107 (1997);
 %%CITATION = PHRVA,D56,3107;%%
  R.~J.~Oakes {\it et al.}, \PRD57, 534 (1998);
 %%CITATION = PHRVA,D57,534;%%
  J.~L.~Feng, J.~F.~Gunion and T.~Han, \PRD58, 071701 (1998);
 %%CITATION = PHRVA,D58,071701;%%
  B.~C.~Allanach, H.~Dreiner, P.~Morawitz, and 
M.~Williams, \PLB420, 307 (1998);
 %%CITATION = PHLTA,B420,307;%%
  S.~Bar-Shalom, G.~Eilam, J.~Wudka and A.~Soni, \PRD59, 035010 (1999);
                                                  \PRL80, 4629 (1999);
 %%CITATION = PHRVA,D59,035010;%%
 %%CITATION = PRLTA,80,4629;%% 
   K.~Hikasa, J.~M.~Yang, and B.-L.~Young, \PRD60, 114041 (1999);
 %%CITATION = PHRVA,D60,114041;%%
    H.~Dreiner, P.~Richardson, and M.~H.~Seymour, 
                     hep-ph/9912407; hep-ph/0007228; 
 %%CITATION = HEP-PH 9912407;%% 
 %%CITATION = HEP-PH 0007228;%% 
    P.~Chiappetta {\it et al.},  \PRD61, 115008 (2000); 
 %%CITATION = PHRVA,D61,115008;%%
    M.~Chemtob, and G.~Moreau, \PRD61, 116004 (2000); \PRD59, 055003 (1999); 
 %%CITATION = PHRVA,D61,116004;%%
 %%CITATION = PHRVA,D59,055003;%%
    T.~Han, and M.~B.~Magro, \PLB476, 79 (2000); 
 %%CITATION = PHLTA,B476,79;%%
    G.~Moreau, E.~Perez, and G.~Polesello, hep-ph/0003012;
 %%CITATION = HEP-PH 0003012;%%
    K.J.~Abraham, K.~Whisnant, J.~M.~Yang, and B.-L.~Young, hep-ph/0007280;
 %%CITATION = HEP-PH 0007280;%%
    G.~Moreau, hep-ph/0012156;  hep-ph/0009140;
 %%CITATION = HEP-PH 0012156;%% 
 %%CITATION = HEP-PH 0009140;%%
    F.~Deliot, G.~Moreau, and C.~Royon, hep-ph/0007288;
 %%CITATION = HEP-PH 0007288;%%
    M.~Chaichian, K.~Huitu, and Z.~H.~Yu, hep-ph/0007220.
 %%CITATION = HEP-PH 0007220;%%
\bibitem{soft_rv} M.~Nowakowski and A.~Pilaftsis, \NPB461, 19 (1996).
 %%CITATION = NUPHA,B461,19;%% 
\bibitem{lynne}
    M.~Bisset, O.~C.~W.~Kong, C.~Macesanu and L.H. Orr, \PLB430, 274 (1998); 
 %%CITATION = PHLTA,B430,274;%%
   C.-H. ~Chang and T.-F. ~Feng, hep-ph/9908295; hep-ph/9901260. 
 %%CITATION = HEP-PH 9908295;%%
 %%CITATION = HEP-PH 9901260;%%
\bibitem{bcs} G.~Passarino and M.~Veltman,  \NPB160, 151 (1979).
 %%CITATION = NUPHA,B160,151;%% 
\bibitem{epsilon} V.~Barger, K.~Cheung, R.J.N. Phillips and 
A. Stange,    \PRD46, 4914 (1992).
 %%CITATION = PHRVA,D46,4941;%%
\bibitem{LEP_higgs} For recent LEP2 limits on the SUSY Higgs sector,
see, {\it e.~g.}, the L3 collaboration, hep-ex/0012017.
 %%CITATION = HEP-EX 0012017;%%
\bibitem{pertur}   B.~Brahmachari and P.~Roy, \PRD50, 39 (1994).
 %%CITATION = PHRVA,D50,39;%%
\bibitem{bsg} See, {\it e.~g.},  
 W. deBoer {\it et al.}, hep-ph/9712376;
 %%CITATION = HEP-PH 9712376;%%
 H. Baer and M. Brhlik, \PRD55, 3201 (1997).
  %%CITATION = PHRVA,D55,3201;%%
\end{thebibliography}
\end{document}